\documentclass[journal]{IEEEtran}
\usepackage{cite}
\usepackage{amsmath,amssymb,amsfonts}
\usepackage{algorithmic}
\usepackage{graphicx}
\usepackage{textcomp}
\usepackage{gensymb}

\usepackage[caption=false,font=footnotesize]{subfig}
\usepackage{multirow}
\usepackage{float}
\usepackage{xcolor}

\begin{document}
%
\title{Energy- and Quality-aware Video Request Policy for Wireless Adaptive Streaming Clients}
%
%
%

\author{C\'esar~D\'iaz, Antonio~Fern\'andez, Fernando Sacrist\'an, and~Narciso~Garc\'ia
\thanks{Manuscript received XXX, 2020; revised YYY, 2020.}
\thanks{C.~D\'iaz, A.~Fern\'andez, and N.~Garc\'ia are with the Grupo de Tratamiento de Im\'agenes, Information Processing and Telecommunications Center and ETSI Telecomunicaci\'on, Universidad Polit\'ecnica de Madrid, 28040 Madrid, Spain (e-mail: cdm@gti.ssr.upm.es, afc@gti.ssr.upm.es, narciso@gti.ssr.upm.es)}
\thanks{F. Sacrist\'an is with Nokia Spain S.A., Mar\'ia Tubau 9,
28050 Madrid, Spain (e-mail: fernando.sacristan\_llorente@nokia.com )}
\thanks{This work has been partially supported by the Ministerio de Ciencia, Innovación y Universidades (AEI/FEDER) of the Spanish Government under project TEC2016-75981 (IVME) and the Spanish Administration agency CDTI under project IDI-20170957 (5GSTB).}}

\markboth{IEEE Transactions on Consumer Electronics,~Vol.~XX, No.~X, October~2020}%
{D\'iaz \MakeLowercase{\textit{et al.}}: Energy- and Quality-aware Video Request Policy for Wireless Adaptive Streaming Clients}

\maketitle

\begin{abstract}
We present a straightforward, non-intrusive adaptive bit rate streaming segment quality selection policy which aims at extending battery lifetime during playback while limiting the impact on the user's quality of experience, thus benefiting consumers of video streaming services. This policy relies on the relationship between the available channel bandwidth and the bit rate of the representations in the quality ladder. It results from the characterization of the energy consumed by smartphones when running adaptive streaming client applications for different network connections (Wifi, 4G, and 5G) and the modeling of the energy consumed as a function of said relationship. Results show that a significant amount of energy can be saved (10 to 30\%) by slightly modifying the default policy at the expense of a controlled reduction of video quality.
\end{abstract}


%
\IEEEpeerreviewmaketitle

\section{Introduction}
\IEEEPARstart{A}{ccording} to Cisco, the annual global IP traffic is expected to reach 4.5 ZB by 2022, and IP video traffic will cover 82\% of it, mostly devoted to over-the-top (OTT) live and on-demand video streaming~\cite{cisco2019cisco}. In addition, the Motion Picture Association of America (MPAA) reported that, during 2019, 75\% of U.S. adults watched movies and television via online subscription services~\cite{mpaa2020comprehensive}. Globally, their popularity has increased to the point that the subscriptions to these platforms’ services outnumbered the rest of the alternatives, including cable. The global number of subscriptions surpassed 863.9 million (an increase of 28\% from the previous year), and its growth is expected to continue.

Among the many consumer gadgets that support OTT media services (handheld devices, desktop and portable computers, streaming sticks, Set-Top Boxes, etc.), smartphones are the most extended autonomous ones for many reasons: mobility, high processing capability, the continuous improvement and expansion of wireless technology and high-speed networks, the massive adoption of HTTP/TCP-based adaptive bit rate (ABR) techniques to stream video content, the proliferation of applications that enable the access to unlimited collections of videos, etc. As of today, there are 3.2 billion connected smartphones in the world~\cite{newzoo2019blobal}, and they are the primary device for viewing video for a 78.4\% of digital video viewers~\cite{doyle2018more}.

However, mobile devices have the same defining disadvantage as any other everyday-use autonomous equipment: a limited battery capacity~\cite{coughlin2015moore}. As of mid-2019, commercial smartphones’ battery capacity can get up to 5000~mAh. For normal use, this represents a battery life of up to 60~hours~\cite{burrell2019best}. However, video applications consume a large amount of energy, draining the battery up to ten times faster. Given the explosion of smartphones for the consumption of audiovisual content via ABR media apps, it seems reasonable to look into how the functioning of the control policy that governs its behavior and, more generally, the session context impact on the consumed energy to extend the battery lifetime.

The sources of energy consumption in mobile video applications are: display, memory usage, decoding, and video-related communications. The energy consumption of the display depends, besides on the hardware involved (e.g. screen size, pixel density...), on the screen brightness, which is adjusted manually or automatically generally considering the environmental lighting, the characteristics of the content, and the user's preferences. The memory usage consumption is proportional to the number of writing and reading operations and so increases with the video bit rate. The impact of media decoding in energy consumption depends on the selected representation video resolution, codec, coding algorithm, and type of decoder used, that is, hardware or software. Finally, wireless communications represent a rather significant fraction of the energy consumption. The longer the receiver is on, the more energy will be consumed.

Many energy-saving or energy-aware strategies have been proposed over the years addressing one or more of the above mentioned factors with the aim of extending the battery lifetime of such an essential consumer device. For the sake of generalization and simplicity, which are key for the scalability of the solution and rapid integration in consumer appliances, we are interested in the ones related to the ABR delivery paradigm that do not interfere with the quality ladder and the integrity of the segments specified in the manuscript, the user-selected smartphone settings, the running mode of the device, or other elements unconnected to the streaming service application itself. That is, this paper focuses on strategies based on performing energy-aware adaptations or revisions of the segment request policy of the client media player control module. However, most of the proposals in the literature were conceived for video delivery paradigms different from the standard ABR technology, were designed to intervene in other aspects of the communications chain, or are based on the modification of the functioning of standard software or protocols. In this sense, many of the works are related specifically to traditional server-driven streaming, where they propose the inclusion of transcoding mechanisms to save energy while preventing quality drops, either based on joint source–channel coding schemes~\cite{wu2020power}, online video framerate variations~\cite{kim2014balancing}, bit rate truncation~\cite{edstrom2019content}, or other techniques aiming at changing the complexity of the streamed video~\cite{mallikarachchi2018decoding},~\cite{munoz2020methodology}. Others are intrusive in the sense that they aim at modifying user-selected running settings like the display brightness~\cite{leu2017energy},~\cite{yan2016rate}, intervene in the device hardware (e.g. CPU frequency~\cite{groba2019qoe},~\cite{tang2017closed}), core software (e.g. OS~\cite{groba2019closed}, or HTTP functioning~\cite{lee2015user}), or in the transmission mean, through a smart allocation of resources (e.g. bandwidth aggregation~\cite{zou2018balancing}, data distribution~\cite{zhao2018performance},~\cite{go2015energy}, or transmission scheduling~\cite{yang2019energy}). Among the ABR-oriented, non-intrusive approaches, some lack completeness, as they do not consider important parameters like the dynamics of the network connection~\cite{petrangeli2016energy}, or present excessively complex models or heuristics to drive the client control policy~\cite{koo2019seamless},~\cite{jo2018video}.

Given this panorama, this work proposes the use of an extremely simple, non-intrusive set of segment request modes that override the decisions of the default policy to boost energy savings of smartphones while impacting the user's experience as little as possible. Its simplicity and the fact that its implementation is not device- or hardware-dependent looks for a rapid implantation in smartphones with the objective of potentially benefiting millions of users. The policy is the outcome of the study of the impact of several variables and the estimation of their weight in the smartphone energy consumption. This has been done through the characterization of three commercial smartphones using three wireless network connections: Wifi, 4G, and 5G. Additionally, we have modeled the behavior of the energy consumed by smartphones versus the available channel bandwidth and the requested representation bit rate, as the relationship between these two turned out to be the most relevant factor affecting smartphone energy consumption. The importance of this parameter was not captured by reference works analyzing the efficiency of handheld devices while performing video delivery from a more general perspective, like that of Trestian et al.~\cite{trestian2012energy}. Finally, based on the resulting model, we propose several energy-saving modes, which represent different trade-offs between saved energy and presented video quality.

The paper is structured as follows. Section~\ref{sec:study_impact_context_energy} describes the study of the impact of the ABR session-related variables on the energy consumed by the handheld device. In Section~\ref{sec:energy_capacity_model}, we include the model that relates the energy consumed with the channel capacity, whereas in Section~\ref{sec:energy_saving_modes}, we present the energy-saving modes derived from that model. Finally, conclusions are included in Section~\ref{sec:conclusion}.

\section{Study of the impact of the context on the energy consumption}
\label{sec:study_impact_context_energy}

We have carried out a study to identify the influence of several high-level variables that are key on the overall energy that the smartphone consumes during ABR sessions. In particular, we have analyzed the impact of the following variables, which are direct or indirect indicators of the video decoding performance, the memory usage and the wireless communications dynamics, on the energy consumption: device, type of network connection, video codec, video resolution and bit rate (size) of the segments.

To perform this analysis, we have conducted two separate sets of experiments. They basically differ in the role given to the network dynamics in the process, represented through the relationship between the available bandwidth and the bit rate of the representations to be downloaded.

So, in the first set, the impact of network conditions is removed by using the maximum possible available bandwidth in the network, which is much higher than the bit rate of any of the available representations. In particular, we use a IEEE 802.11ac Wifi standard-capable router, which enables 'real' connection speeds of up to a few hundreds of Mbps for single client devices. In this way, the time required to download segments belonging to any of the representations is significantly shorter than their length: in the order of tens to hundreds of milliseconds for segments of 6 seconds.

In the second one, we do take into consideration the available bandwidth, which affects the time it takes the client to download the segments: the closer the channel bandwidth and the selected representation bit rate are, the longer it takes to download a segment belonging to that representation.

Nevertheless, in both sets of experiments, we guarantee that the available bandwidth is always high enough to avoid buffer underruns and thus video stalls.

\subsection{Tests not considering network dynamics}
\label{sec:study_no_network_dynamics}
First, we present the features of the first set of experiments. Later on, we present and analyze the results that were obtained.

\subsubsection{Features of the tests}

\paragraph{Content} The content used is the 3840p@24fps 12-minute-long open short movie 'Tears of Steel'~\cite{blender2012tears}, which holds a $~$2.24:1 frame aspect ratio. The original sequence was transcoded (preserving the framerate) and segmented for ABR streaming (MPEG-DASH~\cite{dash2019dynamic}), obtaining 8 representations whose characteristics are detailed in Table~\ref{tab:energy_duration}.

\begin{table}[h!]
\renewcommand{\arraystretch}{1.3}
  \centering
  \caption{\label{tab:energy_duration}Representations Used in the First Set of Tests.}
  \vspace{-1mm}
    \begin{tabular}{cccc}
      \hline \hline
      \multirow{2}{*}{\textbf{Codec}} & \multirow{2}{*}{\textbf{Resolution}} & \textbf{Equivalent} & \textbf{bit rate preliminary}\\
       & & \textbf{Resolution} & \textbf{tests [Mbps]}\\
      \hline
      \multirow{4}{*}{AVC} & 428x182 & 240p & 0.8\\
        & 854x382 & 480p & 3\\
        & 1280x572 & 720p & 8\\
        & 1920x858 & 1080p & 20\\
      \hline
      \multirow{4}{*}{HEVC} & 428x182 & 240p & 0.6\\
        & 854x382 & 480p & 2\\
        & 1280x572 & 720p & 4\\
        & 1920x858 & 1080p & 10\\
      \hline\hline
    \end{tabular}
  \vspace{-2mm}
\end{table}

\paragraph{Setup and equipment} The setup consisted of three smartphones, SP A, SP B, and SP C, connected to the internet through an IEEE 802.11ac WiFi connection, an ABR player, and an ABR server. The features of the smartphones are presented in Table~\ref{tab:smartphones_features}.

\begin{table}[h!]
\renewcommand{\arraystretch}{1.3}
\centering
 \caption{\label{tab:smartphones_features}Smartphone Specifications.}
 \vspace{-1mm}
\begin{tabular}{cccc}
    \hline\hline
  \textbf{Feature} & \textbf{SP A} & \textbf{SP B} & \textbf{SP C}\\
  \hline
  \multirow{3}{*}{Processor} & \multirow{3}{*}{8x2.6 GHz} & \multirow{3}{*}{8x1.6 GHz} & 1x2.84 GHz\\
  & & & + 4x1.78 GHz\\
  & & & + 3x2.42 GHz \\
  Battery capacity  & 3000 mAh & 2350 mAh & 3800 mAh\\
  Resolution  & 1440x2560 & 720x1280 & 1080x2340\\
  Screen size & 5.1" & 4.7" & 6.39"\\
  Pixel density & 577 ppi & 312 ppi & 403 ppi\\
  Hardware video & \multirow{2}{*}{AVC/HEVC} & \multirow{2}{*}{AVC/HEVC} & \multirow{2}{*}{AVC/HEVC}\\
  decoding & & &\\
  \hline\hline
\end{tabular}
\vspace{-4mm}
\end{table}

The features of the setup were the following:
\begin{itemize}
    \item The energy measurement data was gathered via the smartphone's battery fuel gauge and an energy profiling software, which allowed for a fine-enough estimation of the energy consumed during the session. In particular, the state of charge (SoC) of the battery, which only provided integer values, was sampled every 30 seconds.
    \item No other apps run during the tests, besides the ones required to make the device work in default mode.
    \item The reception virtual buffer size was set to 96~MB. Moreover, new segments were requested only whenever the buffer level was below 90\% to avoid downloading segments in bursts and so spread download intervals uniformly over the session.
    \item The devices were placed near a Wifi access point, with no obstacles between the device and the access point, to guarantee radio signal power stability and so maximum bandwidth during the tests.
    \item The screen brightness was set at 90\% for both devices and the speakers were turned off.
\end{itemize}

\paragraph{Procedure} We set up several sessions. During each session, one of the 8 representations of the content was requested and played on a continuous loop until the whole battery drained and the smartphone shut off. Prior to each experiment, the smartphone battery was fully charged and then disconnected from any energy source.

\begin{figure}
    \centering
    \includegraphics[width=.83\columnwidth]{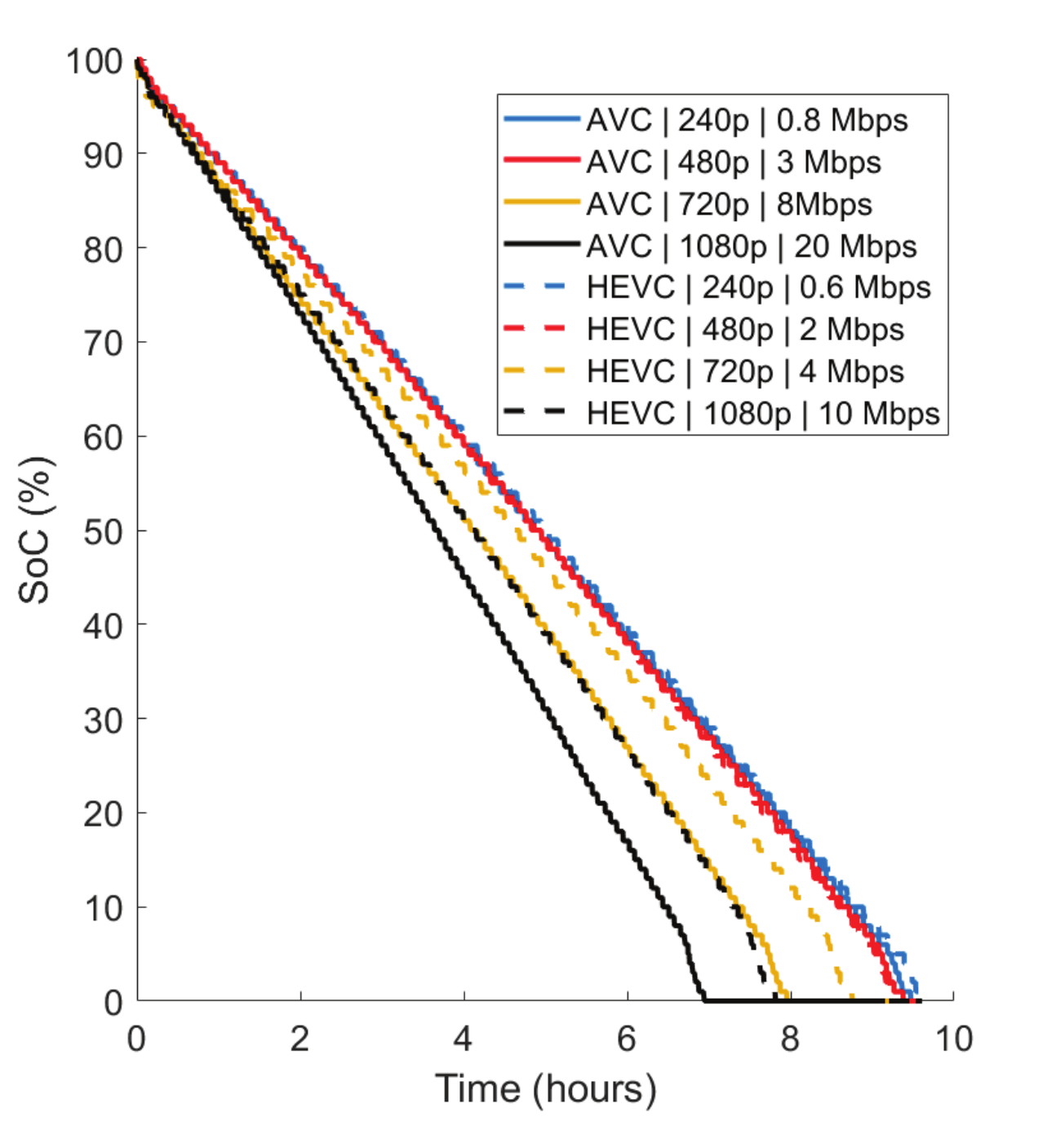}
    \vspace{-4mm}
    \caption{Battery drainage of SP B for different representations}
    \label{fig:battery_duration}
    \vspace{-6mm}
\end{figure}

\paragraph{Results} Fig.~\ref{fig:battery_duration} shows the evolution of the SoC over time during the continuous ABR session for smartphone B for the different representations. The results of the rest of the devices present analogous behaviors and thus are not provided to save space. We can see that the evolution of the SoC follows a fairly linear behavior throughout the session for all representations. The slope of the line depends on the values of the variables under study. In particular, the energy consumption seems to be highly correlated with the representation bit rate.

Given that the device and the connection are fixed and also that all smartphones are capable of hardware-decoding both AVC and HEVC, the differences between representations with the same resolution and different codec are necessarily due to the segment size. That is, data transmission has a much greater impact on energy consumption than the rest of the tested features, a statement that agrees with previous studies~\cite{hoque2013dissecting}. Furthermore, for perceptually equivalent representations, HEVC should be chosen over AVC to reduce energy consumption, due its greater coding efficiency.

Additionally, considering the results, to minimize energy consumption, the lowest quality should always be selected. However, evidently, that would come at the cost of a significant drop in Quality of Experience (QoE).

\subsection{Tests considering network dynamics}
\label{sec:energy_consumption}
In this set of tests, in light of the previous conclusions, we have thoroughly analyzed the influence of the available bandwidth, $BW$ and the representation bit rate, $B_{\text{rep}}$, on the energy consumed by the smartphone during an ABR session. In particular, we have focused the experiments on studying the influence of the relationship between bandwidth and bit rate, that we have called relative bandwidth ($\widetilde{BW}$). As mentioned before, in these tests, we assume that the ABR algorithm always requests and downloads representations whose associated bit rate is below the available channel bandwidth. Therefore:

\begin{equation}
    \widetilde{BW} = \frac{BW}{B_{\text{rep}}} \geq 1
\end{equation}

Moreover, in the same way as for the bandwidth, we define the relative energy consumption of a target representation ($\widetilde{EC}$), for a given combination of factors (device, network connection, and video codec) as:

\begin{equation}
    \widetilde{EC} = \frac{EC_{\text{rep}}}{EC_{\text{ref}}}
\end{equation}
where $EC_{\text{rep}}$ is the average consumption for the representation being tested, and $EC_{\text{ref}}$ is the average energy consumed for the representation with the lowest resolution and bit rate (240p at 0.66 Mbps), which is used as reference. Both values result from averaging all the measurements taken using the same combination of factors (device, network connection, and video codec) and for the above-mentioned representation. In this way, we minimize any fuel gauge measurement errors. The reference representation resolution and bit rate were selected because they are low enough to minimize computational cost and network load, yet they still allow us to take into account screen energy expenditure.

Since our ultimate goal is to minimize the energy consumption while maximizing the QoE for some base conditions (that is, for a fixed combination of device, network and codec), $\widetilde{EC}$ can expose the dependency between the representation and the channel bandwidth, regardless of the base conditions, thus enabling a proper overall comparison.

\subsubsection{Features of the tests}

\paragraph{Content}
We also used 'Tears of Steel' as source content. Again, the original sequence was transcoded (preserving the framerate) and segmented for ABR streaming, obtaining in this case 8 representations whose characteristics are detailed in Table~\ref{tab:representations_tests_network_dynamics}. The target encoding bit rates for every resolution are 0.66, 5, 10, 20 and 30 Mbps.

\begin{table}[h!]
\renewcommand{\arraystretch}{1.3}
\vspace{-3mm}
  \begin{center}
  \caption{\label{tab:representations_tests_network_dynamics}Representations Used in the Second Set of Tests.}
  \vspace{-1mm}
    \begin{tabular}{cccc}
      \hline\hline
      \multirow{2}{*}{\textbf{Codec}} & \multirow{2}{*}{\textbf{Resolution}} & \textbf{Equivalent} & \textbf{bit rate preliminary}\\
       & & \textbf{Resolution} & \textbf{tests [Mbps]}\\
      \hline
      \multirow{4}{*}{AVC} & 428x182 & 240p & \multirow{4}{*}{0.66, 5, 10, 20, 30}\\
        & 854x382 & 480p & \\
        & 1280x572 & 720p & \\
        & 1920x858 & 1080p & \\
      \hline
      \multirow{4}{*}{HEVC} & 428x182 & 240p & \multirow{4}{*}{0.66, 5, 10, 20, 30}\\
        & 854x382 & 480p & \\
        & 1280x572 & 720p & \\
        & 1920x858 & 1080p & \\
        \hline\hline
    \end{tabular}
  \end{center}
  \vspace{-3mm}
\end{table}

\paragraph{Setup and equipment}
It consisted of the same three smartphones (see Table~\ref{tab:smartphones_features}) connected to the internet through a wireless connection (IEEE 802.11ac WiFi, 4G, and 5G NSA -non-standalone mode-), an ABR player, and an ABR server that can limit content delivery speed by request. Specifically, we used the combinations of smartphones and connections depicted in Table~\ref{tab:smartphones_connections}.

\begin{table}[h!]
\renewcommand{\arraystretch}{1.3}
\vspace{-3mm}
\centering
\caption{\label{tab:smartphones_connections}Combinations of Smartphones and Connections.}
\vspace{-1mm}
\begin{tabular}{cccc}
\hline\hline
  & \textbf{Smartphone A} & \textbf{Smartphone B} & \textbf{Smartphone C}\\
  \hline
  Wifi & Yes & Yes & Yes\\
  4G  & No & No & Yes\\
  5G  & No & No & Yes\\
  \hline\hline
\end{tabular}
\end{table}

The Wifi and 4G connections were used with a bandwidth limiter software to be able to set specific $\widetilde{BW}$ values. In particular, for every representation, we considered two kinds of measurements. In the first one, the connections were not controlled or limited artificially, but the bandwidth fluctuated freely. In this way, we were able to generate easily many $\widetilde{BW}$ points within a wide range of values. In the second scenario, the bandwidth was set to the bit rate of the representation, that is, $\widetilde{BW} \approx 1$. So, we were able to obtain relative bandwidth points close to the minimum limit. Finally, only the first scenario was implemented for the 5G connection.

A broader scenario was considered for the energy consumption test to gather an ample set of measures. Thus, different device locations for different sessions were used in indoor measurements involving a WiFi connection to be able to obtain abundant $\widetilde{BW}$ points in the whole range. On the other hand, for the 4G-related tests, the device was in an outdoor location to guarantee signal coverage, while trying to achieve the same diversity of conditions than in the WiFi connections. Finally, for the 5G-related tests, the device was located outdoors and very close to one of the 5G antennas available at the time of the experimentation to guarantee signal reception.

\paragraph{Procedure}
We set up a large number of sessions. For each session, one of the 8 representations of the content was requested and played, resulting in a new measurement. It is important to note that the average current values referred here include only the energy consumed by the smartphone on account of the ABR session. Any other energy spent by the rest of the processes is not considered in these figures.

\paragraph{Results}

\begin{figure*}
    \centering
    \subfloat[\label{fig:avg_current_vs_avg_bw}Absolute values ($EC$ vs $BW$)]{\includegraphics[width=.48\linewidth]{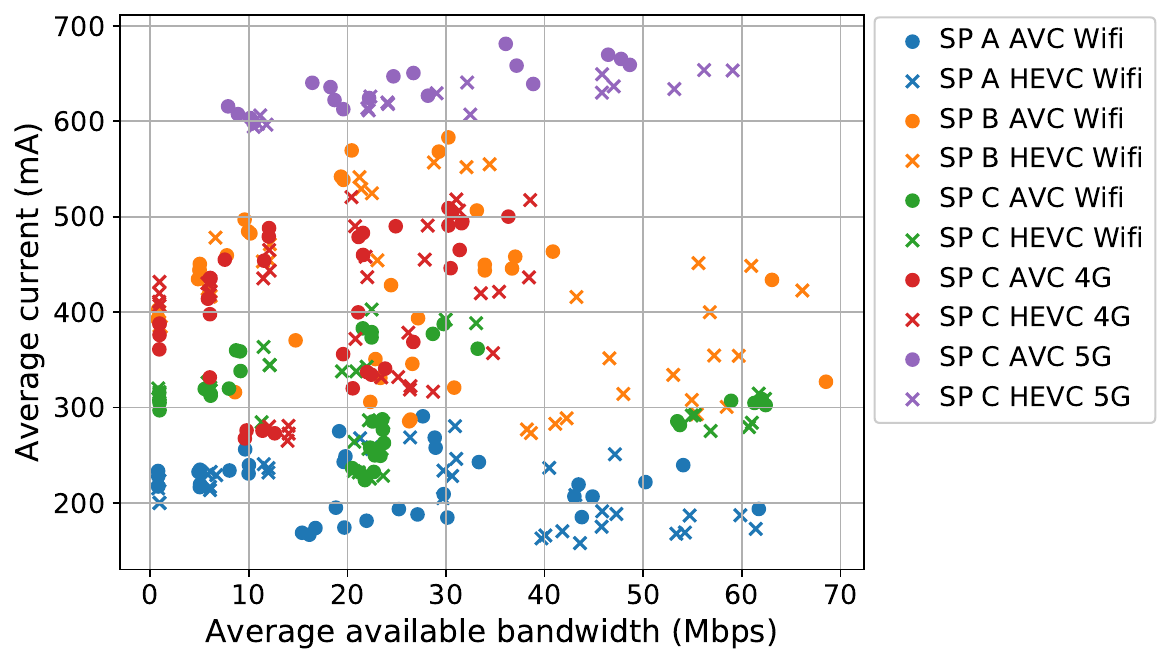}}
    \subfloat[\label{fig:rel_avg_current_vs_rel_avg_bw}Relative values ($\widetilde{EC}$ vs $\widetilde{BW}$)]{\includegraphics[width=.48\linewidth]{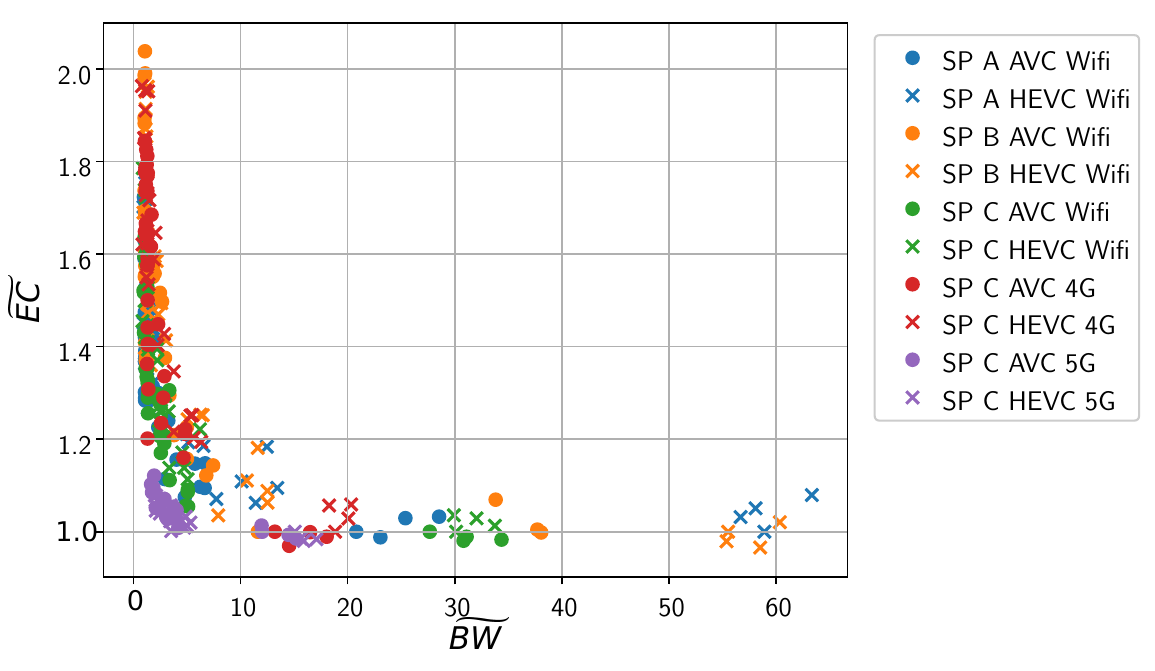}}
    \caption{\label{fig:current_vs_bw}Average current consumed vs average available bandwidth}
    \vspace{-0.5cm}
\end{figure*}

Figure~\ref{fig:avg_current_vs_avg_bw} depicts the impact of the available channel bandwidth on the average current (in mA) for all the combinations of device, connection, resolution, codec and bit rate considered in the tests. 
Every color includes the results for all the combinations sharing device, connection and codec. As can be observed, there is a strong correlation between representation bit rate, device, type of connection, and the energy consumed by the device. Indeed, it can be seen that the device hardware is decisive in the energy it consumes. Moreover, regarding device and connection, the one consuming the lowest on average is the SP A using Wifi, followed by SP C using Wifi. Then comes SP B using Wifi and, closely, SP C using 4G. Finally, on the top of energy expenditure we find the SP C using 5G. Thus, using the SP C 5G radio access to download segments seems to be energetically inefficient, specially having also the 4G radio access, which consumes barely half the energy. The reason to that is that the discontinuous reception (DRX) mode is not implemented in 5G and so the receiver is never turned off, even if there is no data to transfer. Therefore, lacking an implementation of this reception mode, we advise in favor of using 4G or Wifi when available to save energy.

The figure also shows that the results follow no clear patterns with respect to the absolute values for channel bandwidth. Hence, the latter are not reliable enough sources to properly evaluate the impact of the network dynamics on the energy that the device uses under a given combination of elements (device, codec, connection). Therefore, we use the relative ones defined above: $\widetilde{BW}$ and $\widetilde{EC}$. Moreover, using relative values is also more interesting for determining a representation selection policy that minimizes energy consumption while maximizing visual quality, for a specific ABR session.

So, in Figure~\ref{fig:rel_avg_current_vs_rel_avg_bw}, we can find the same data as in Figure~\ref{fig:avg_current_vs_avg_bw}, however this time expressed in terms of $\widetilde{BW}$ and $\widetilde{EC}$. Using this new format, patterns become clearer, enabling a proper analysis and the possibility to draw robust conclusions.

In particular, the figure shows that the energy consumed by the device to download, decode, and present some content decreases exponentially with the relationship between the bandwidth available in the channel and the bit rate of the representation that is requested. Indeed, the more available bandwidth in relation to the bit rate of the segments, the less time it takes the client to obtain those segments, and therefore the less energy is consumed. On the other hand, if $\widetilde{BW}$ gets closer to 1, the energy consumed can be between 50\% and 100\% higher when compared with the minimum average current for the codec, network and device. Therefore, $\widetilde{BW}$ is key in determining the impact of the ABR session in the energy consumed by the device.

In addition, observing the previous figures, several dynamics can be pointed out. First, if the bit rate is fixed for a given combination of device, connection, and codec, the impact of resolution is negligible on energy consumption. Therefore, selecting a version of the video with a lower resolution will not help the implementation of a energy-saving mode. Second, if the video resolution is fixed for a given combination of device, connection, and codec, the impact of bit rate on energy consumption is quite noticeable. The greater the size of the segments downloaded (better representation), the greater the average consumption. Therefore, selecting a version of the video with a lower bit rate does make sense as part of the strategy to save energy. Furthermore, this action would result in increasing the relative bandwidth, hence decreasing further the average consumption of the device during the ABR session. Third, for a given combination of device, connection, resolution and bit rate, the codec that is used does not affect much the energy spent by the smartphone. This result confirms that it is always better to select HEVC over AVC, as it compresses content more efficiently, which translates into smaller segments and so into energy savings.

In light of the experiments, in order to reduce energy consumption while impacting as little as possible in the user’s QoE, it is advisable neither to select representations whose bit rate is too low nor  too close to the available bandwidth. The operation points in between represent trade-off points between QoE and battery life to be considered. The exact points to target will be defined by the different energy-saving modes. To properly define these modes, we have first modeled the relative energy consumed by a smartphone as a function of the relative bandwidth. This is described in the following section.

\begin{figure*}
    \centering
    \includegraphics[width=.97\linewidth]{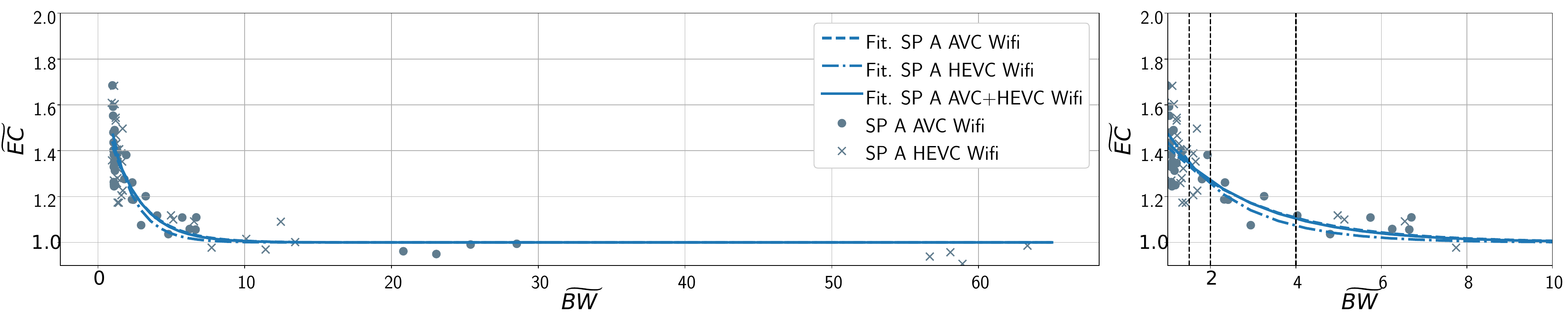}\\
    \includegraphics[width=.97\linewidth]{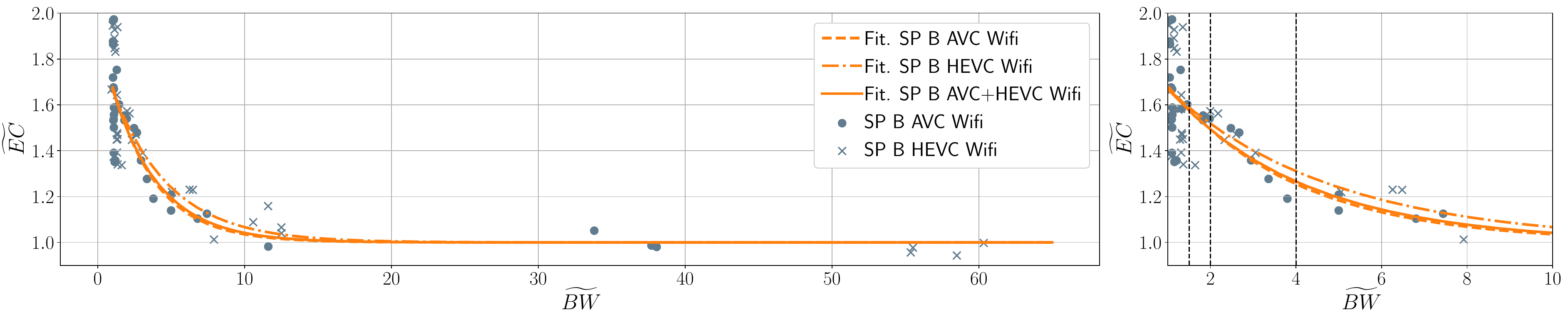}\\
    \includegraphics[width=.97\linewidth]{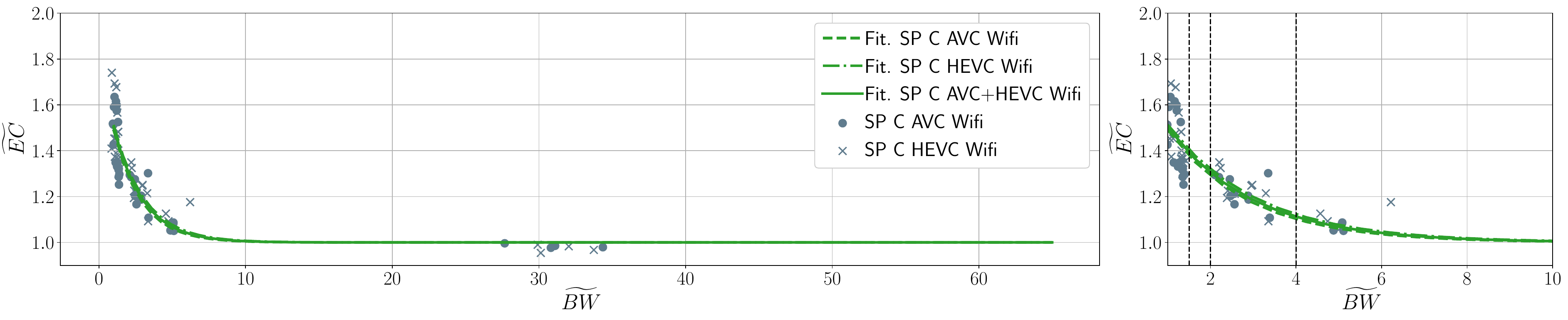}\\
    \includegraphics[width=.97\linewidth]{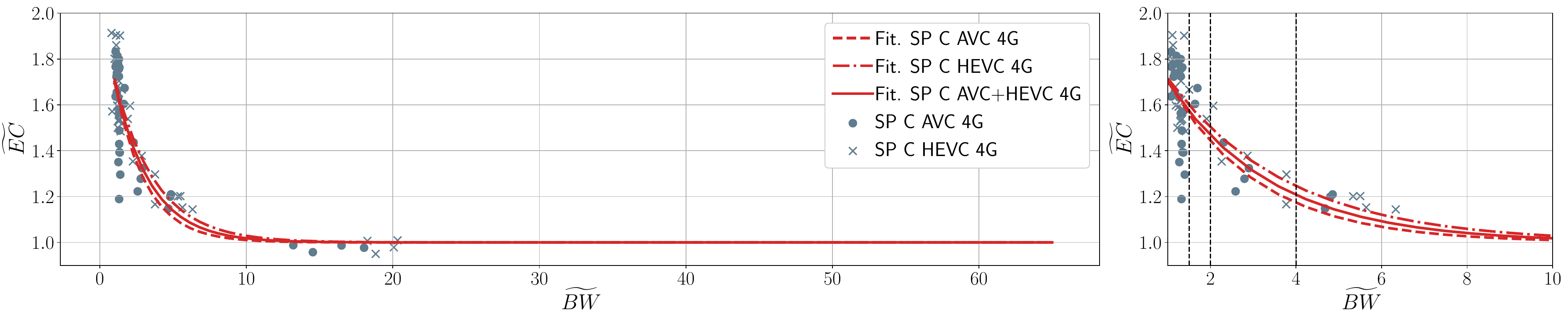}\\
    \includegraphics[width=.97\linewidth]{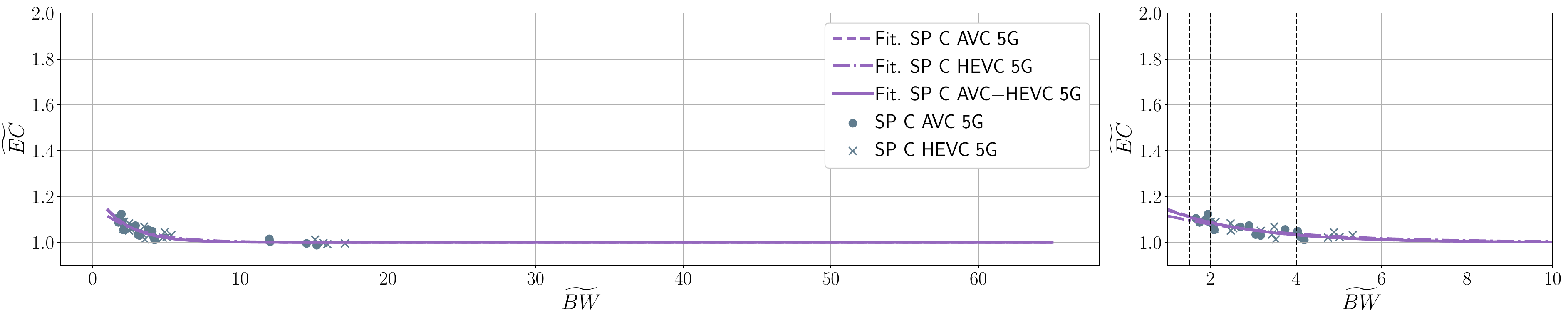}\\
    \includegraphics[width=.97\linewidth]{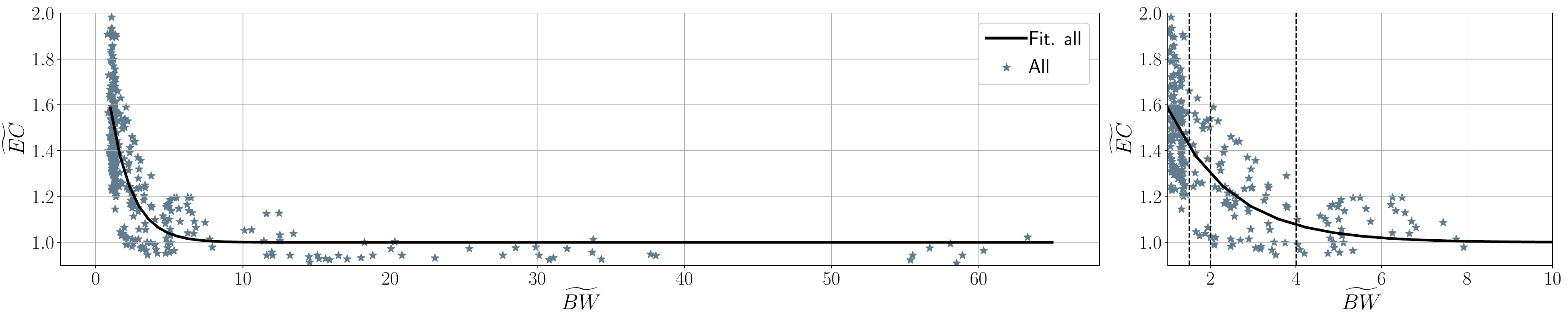}\\
    \caption{\label{fig:fit_rel_avg_current_vs_rel_avg_bw}Left: curve fitting of the average relative current consumed ($\widetilde{EC}$) vs average relative available bandwidth ($\widetilde{BW}$) for different combinations of device, network connection and codec. The
bottom-most curve includes all the data combined. Right: zoom of the same graphs for the range of values of $\widetilde{BW}$ between 1 and 10, which is the main area of interest.}
\end{figure*}

\begin{table*}[h!]
\renewcommand{\arraystretch}{1.3}
\caption{\label{tab:fit_rel_avg_current_vs_rel_avg_bw}Fitting Curve Values ($a$, $b$, and $c$), and Correlation Coefficients ($R^2$, PCC, and SROCC) for All Combinations of Smartphone, Network Connection and Codec, and for All the Data Points Considered Simultaneously (Overall Results).}
\vspace{-4mm}
  \begin{center}
    \begin{tabular}{ccccccccc}
      \hline\hline
      \multirow{2}{*}{\textbf{Smartphone}} & \textbf{Network} &
      \multirow{2}{*}{\textbf{Codec}} & \multicolumn{3}{c}{\textbf{Fitting curve}} & \multirow{2}{*}{\textbf{$R^2$}} &
      \multirow{2}{*}{\textbf{PCC}} &
      \multirow{2}{*}{\textbf{SROCC}}\\
      \cline{4-6}
       & \textbf{connection} & & \textbf{$a$} & \textbf{$b$} & \textbf{$c$} & & &\\
      \hline
      \multirow{3}{*}{SP A} & \multirow{3}{*}{Wifi (IEEE 802.11ac)} & AVC & 0.653 & 0.452 & 1.000 & 0.759 & 0.871 & 0.883\\
      & & HEVC & 0.890 & 0.628 & 1.000 & 0.754 & 0.868 & 0.834\\
      & & AVC+HEVC & 0.704 & 0.480 & 1.000 & 0.724 & 0.851 & 0.803\\
     \hline
      \multirow{3}{*}{SP B} & \multirow{3}{*}{Wifi (IEEE 802.11ac)} & AVC & 0.947 & 0.329 & 1.000 & 0.711 & 0.843 & 0.830\\
      & & HEVC & 0.863 & 0.256 & 1.000 & 0.744 & 0.863 & 0.863\\
      & & AVC+HEVC & 0.911 & 0.308 & 1.000 & 0.728 & 0.853 & 0.826\\
      \hline
      \multirow{9}{*}{SP C} & \multirow{3}{*}{Wifi (IEEE 802.11ac)} & AVC & 0.828 & 0.524 & 1.000 & 0.804 & 0.897 & 0.942\\
      & & HEVC & 0.825 & 0.476 & 1.000 & 0.811 & 0.900 & 0.937\\
      & & AVC+HEVC & 0.826 & 0.499 & 1.000 & 0.786 & 0.887 & 0.919\\
      \cline{2-9}
       & \multirow{3}{*}{4G} & AVC & 1.121 & 0.468 & 1.000 & 0.721 & 0.849 & 0.820\\
      & & HEVC & 1.021 & 0.356 & 1.000 & 0.852 & 0.923 & 0.816\\
      & & AVC+HEVC & 1.051 & 0.406 & 1.000 & 0.769 & 0.877 & 0.826\\
      \cline{2-9}
       & \multirow{3}{*}{5G (NSA)} & AVC & 0.238 & 0.500 & 1.000 & 0.832 & 0.912 & 0.948\\
      & & HEVC & 0.167 & 0.373 & 1.000 & 0.791 & 0.889 & 0.903\\
      & & AVC+HEVC & 0.229 & 0.489 & 1.000 & 0.807 & 0.898 & 0.915\\
      \hline
      \multicolumn{3}{c}{Overall} & 1.154 & 0.677 & 1.000 & 0.665 & 0.815 & 0.828\\
    \hline\hline
    \end{tabular}
  \end{center}
  \vspace{-6mm}
\end{table*}

\section{Energy consumption vs channel capacity model}
\label{sec:energy_capacity_model}
We have modeled the average relative energy consumed by the smartphones as a function of the relative bandwidth to propose more suitable energy-saving modes and evaluate their outcomes more precisely. To that end, we have used an exponential function to fit the available data:
\begin{equation}
    \widetilde{EC} = a \cdot \text{exp}(-b \cdot \widetilde{BW}) + c
\end{equation}
For practical reasons, this curve is later on shifted to ensure that the asymptote is $\widetilde{EC}$ = 1. Thus, $c$ = 1.

Figure~\ref{fig:fit_rel_avg_current_vs_rel_avg_bw} shows the different combinations of device, network connection and codec considered in this work. The bottom-most curve includes the exponential fitting curve considering the whole collection of data points, regardless of the combination, to be used as the basis of the proposed energy-saving modes. For all combinations, there is a zoom of the graphs for the range of values of $\widetilde{BW}$ between 1 and 10 is also included, since this is the main area of interest. All the graphs show the same behavior, regardless of the combination. They only differ somewhat in the specific values of the consumed energy for given relative bandwidth points. Hence, the same type of curve can be used for all of them, only changing slightly the values of the curve. The models emphasize that if the available bandwidth is scarce, yet enough compared to the bit rate of the requested quality, the mobile device can consume more that 50\% more energy compared to the situation where the bandwidth is several times greater than the requested bit rate. In the other end, $\widetilde{EC}$ values are below 1.1 (roughly the same level of consumption) for $\widetilde{BW}$ is higher than a point between 4 and 7, approximately, depending on the combination. Although there are some variations regarding the exact numbers, this general tendency is valid for any combination of device and network. Only for the 5G-related data points the tendency is somewhat different. The lack of DRX in the receiver for this type of wireless connection ostensibly reduces the differences between measurements along the x-axis, visibly flattening the associated fitting curve and separating these points from the overall one.

In addition, Table~\ref{tab:fit_rel_avg_current_vs_rel_avg_bw} presents the values of $a$, $b$ and $c$ for each curve, as well as several coefficients that show the correlation between the data points and the curve: the coefficient of determination ($R^2$), the Pearson correlation coefficient (PCC), and the Spearman's rank correlation coefficient (SROCC). As can be seen, the value of the correlation coefficients vary slightly from combination to combination, showing that the exponential curves fit the data points better in some cases than in others. However, the values are high for all the combinations and the case that considers all the data points, which is used as a reference for the energy-saving modes proposed next. 

\section{Energy-saving modes}
\label{sec:energy_saving_modes}

\subsection{Proposed modes}
Based on the previous model and results, we propose that the client selects the representation of the segment to be downloaded considering the following definitions and expressions. Let $R_{\text{rep}} = {B_1,\dots,B_{N_{rep}}}$ be the set of representation bit rates in the quality ladder, where $B_1$ is the bit rate of the lowest quality representation and $B_{N_{rep}}$ is the bit rate of the highest quality representation. Also, let $S_{\text{rep}}$ be the subset of $R_{\text{rep}}$ including all the bit rates lower than or equal to $BW / \gamma$, where $\gamma$ is used to determine the intensity of the selected policy to save energy. Thus, we define 4 modes for $\gamma$:
\begin{itemize}
    \item Light: $\gamma = 1.5$. It aims for savings of around 10\%.
    \item Medium: $\gamma = 2$. It aims for savings of around 20\%.
    \item Strict: $\gamma = 4$. It aims for savings of around 30\%.
    \item Adaptive: it switches from mode Light to Medium and from Medium to Strict depending on the battery SoC.
\end{itemize}
Thus, $\gamma$ is a control parameter which can be used by designers to create new energy-saving modes that more closely fit the requirements or objectives of a specific ABR implementation.

The dashed lines in the zoomed regions in Figure~\ref{fig:fit_rel_avg_current_vs_rel_avg_bw} represent the first three modes. For the sake of comparison, the greatest energy saving that could be achieved given the results and model is close to 40\%, but the video quality that could be offered would be too low, and so it is not considered.

Finally, the bit rate that is selected, $B_{\text{rep}}$ is:
\begin{equation}
    B_{\text{rep}} = \begin{cases}
                            \max{B_i \in S_{\text{rep}}}\text{, if }S_{\text{rep}} \neq \emptyset\\
                            B_1\text{, if }S_{\text{rep}} = \emptyset
            \end{cases}
\end{equation}
That is, the selected representation is the one with the maximum bit rate that meets the condition imposed by the mode. If the channel capacity is particularly low, the mode is rather strict, and the quality ladder is not well defined
, $S_{\text{rep}}$ can contain very few representations or even be empty (in that case, the lowest bit rate in $R_{\text{rep}}$ is selected). Under these circumstances, the effectiveness of the mode might be compromised, resulting in no significant energy savings.

\begin{table}[h!]
\renewcommand{\arraystretch}{1.3}
\vspace{-3mm}
  \begin{center}
  \caption{\label{tab:quality_ladder_experiments}Representations (HEVC-encoded) Used in the Experiments.}
    \begin{tabular}{ccc}
      \hline\hline
      \multirow{2}{*}{\textbf{Resolution}} & \textbf{Equivalent} & \textbf{bit rate}\\
      & \textbf{Resolution} & \textbf{[Mbps]}\\
      \hline
      428x182 & 240p & 0.65\\
      854x382 & 480p & 1.25\\
      1024x458 & 576p & 2.0\\
      1280x572 & 720p & 2.5\\
      1440x644 & 960p & 3.5\\
      1920x858 & 1080p & 5\\
      2560x1144 & 1200p & 7.5\\
      2880x1286 & 1440p & 10\\
      3440x1536 & 1600p & 15\\
      3840x1714 & 2160p & 20\\
      \hline\hline
    \end{tabular}
  \end{center}
  \vspace{-7mm}
\end{table}

\begin{figure*}
    \centering
    \subfloat[\label{fig:rep_energy_strategy}Requested representation]{\includegraphics[width=.64\linewidth]{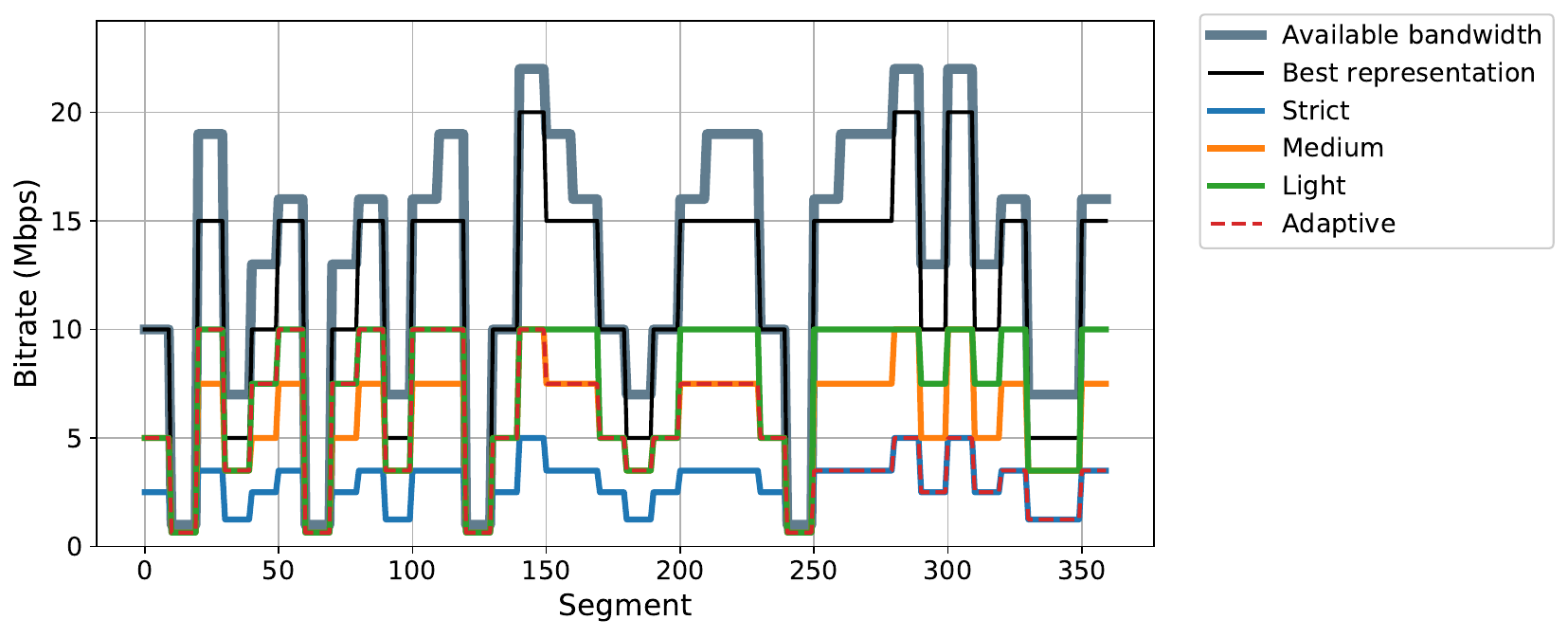}}
    \subfloat[\label{fig:energy_strategy}Energy consumed]{\includegraphics[width=.35\linewidth]{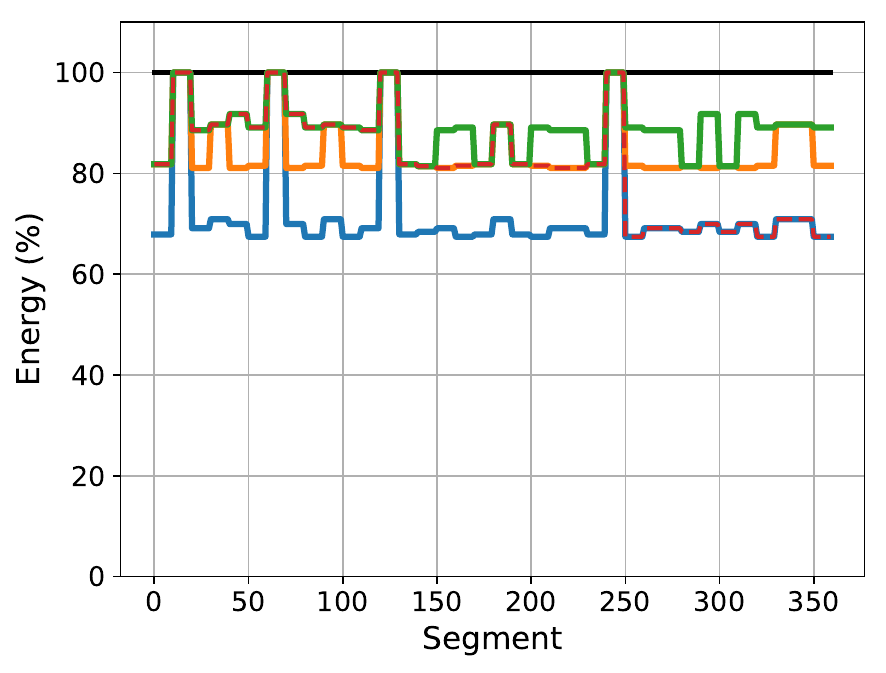}}
    \caption{\label{fig:rep_energy_strategy}(a) Requested representation and (b) energy consumed as a percentage of the one consumed when the best possible representation is requested (energy-saving mode off)
    . In this example, the channel bandwidth varies randomly every minute and remains steady between changes.}
    \vspace{-5mm}
\end{figure*}

\subsection{Implementation}
\subsubsection{Features of  the experiments}
We have tested the four proposed energy-saving modes and compared them to one another and to the case where the energy-saving mode is off (it is either not implemented or not switched on). The results are provided in terms of the average energy consumed and the average image quality shown with respect to the case where the energy-saving mode is off and so the representation whose bit rate is closest to the available bandwidth is the one requested. The former is presented as the percentage of the energy consumed if the energy-mode is off. The latter is presented as the difference between the image quality provided if the energy-mode is off and the one obtained selecting each of the modes. This difference is measured by means of three very popular image quality metrics: PSNR, SSIM~\cite{wang2004image}, and VMAF~\cite{vmafblog} (v0.6.1 applying phone model~\cite{vmafgithub}).

The modes were tested simulating five channel behaviors. A channel behavior is defined as a set of values indicating the capacity of the channel (i.e. the available bandwidth) during consecutive periods of time equal to the length of the segments (6~seconds). To generate the behaviors, the channel capacity was set to take one of the following values during each period: 1, 4, 7, 10, 13, 16, 19, and 22~Mbps. The channel behaviors used in the experiments were the following:
\begin{itemize}
    \item High capacity channel: constant $BW$ = 22 Mbps.
    \item Medium capacity channel: constant $BW$ = 13 Mbps.
    \item Low capacity channel: constant $BW$ = 4 Mbps.
    \item Staircase capacity channel: the available bandwidth is increased in 3~Mbps from one period of time to the next until it reaches the maximum. Afterwards, the capacity is decreases in the same amount until it reaches the minimum. Then, it starts over.
    \item Random capacity channel: the available bandwidth is selected randomly out of the available values every ten periods of time. During these periods, it remains constant.
\end{itemize}

\begin{table*}
\renewcommand{\arraystretch}{1.3}
  \begin{center}
  \caption{\label{tab:impact_strategies_on_quality_energy}Impact of the Different Strategies on the Average Image Quality Presented to the User (Measured via PSNR, SSIM and VMAF) and the Average Energy Consumed as a Percentage of the One Consumed if the Best Representation is Always Selected. Both the Absolute and Differential Values with Respect to the Ones Obtained when the Energy-saving Mode is Off are Presented.}
  \vspace{-1mm}
    \begin{tabular}{ccccccccc}
      \hline\hline
      \textbf{Channel} & \textbf{energy-saving mode} &
      \textbf{Energy consumed (\%)} &
      \textbf{PSNR} &
      \textbf{$\Delta$PSNR} &
      \textbf{SSIM} &
      \textbf{$\Delta$SSIM} &
      \textbf{VMAF} &
      \textbf{$\Delta$VMAF}\\
      \hline
      \multirow{5}{*}{High capacity} & Off & 100.0 & 43.2 & - & 0.9987 & - & 100.0 & -\\
      & Light & 81.42 & 41.7 & 1.5 & 0.9976 & 0.0011 & 99.79 & 0.21\\
      & Medium & 81.42 & 41.7 & 1.5 & 0.9976 & 0.0011 & 99.79 & 0.21\\
      & Strict & 68.40 & 39.9 & 3.3 & 0.9946 & 0.0041 & 97.02 & 2.98\\
      & Adaptive & 77.08 & 41.0 & 2.2 & 0.9966 & 0.0021 & 98.87 & 1.13 \\
     \hline
      \multirow{5}{*}{Medium capacity} & Off & 100.0 & 41.7 & - & 0.9976 & - & 99.79 & -\\
      & Light & 91.77 & 41.0 & 0.7 & 0.9967 & 0.0009 & 99.10 & 0.70\\
      & Medium & 81.06 & 39.9 & 1.8 & 0.9946 & 0.0030 & 97.02 & 2.77\\
      & Strict & 69.94 & 37.5 & 4.2 & 0.9857 & 0.0119 & 88.80 & 10.99\\
      & Adaptive & 77.35 & 39.0 & 2.8 & 0.9916 & 0.0060 & 94.28 & 5.51\\
      \hline
      \multirow{5}{*}{Low capacity} & Off & 100.0 & 38.8 & - & 0.9912 & - & 93.72 & -\\
      & Light & 90.75 & 37.5 & 1.3 & 0.9857 & 0.0055 & 88.80 & 4.91\\
      & Medium & 81.42 & 34.5 & 3.0 & 0.9617 & 0.0240 & 72.09 & 16.71\\
      & Strict & 73.20 & 31.6 & 5.9 & 0.9150 & 0.0706 & 51.51 & 37.29\\
      & Adaptive & 84.87 & 33.9 & 3.6 & 0.9541 & 0.0316 & 70.80 & 18.00\\
      \hline
      \multirow{5}{*}{Staircase} & Off & 100.0 & 38.2 & - & 0.9864 & - & 92.73 & -\\
      & Light & 89.14 & 37.6 & 0.6 & 0.9845 & 0.0019 & 91.19 & 1.54\\
      & Medium & 85.16 & 37.3 & 0.8 & 0.9833 & 0.0031 & 90.20 & 2.52\\
      & Strict & 72.51 & 35.1 & 3.0 & 0.9675 & 0.0189 & 79.65 & 13.08\\
      & Adaptive & 82.23 & 36.5 & 1.6 & 0.9784 & 0.0081 & 86.96 & 5.76\\
      \hline
      \multirow{5}{*}{Random} & Off & 100.0 & 38.7 & - & 0.9884 & - & 94.15 & -\\
      & Light & 88.93 & 38.2 & 0.5 & 0.9870 & 0.0014 & 93.12 & 1.02\\
      & Medium & 84.61 & 38.0 & 0.7 & .9864 & 0.0020 & 92.60 & 1.55\\
      & Strict & 72.28 & 36.3 & 2.4 & 0.9775 & 0.0109 & 85.07 & 9.08\\
      & Adaptive & 82.02 & 37.4 & 1.4 & 0.9836 & 0.0048 & 90.25 & 3.89\\
      \hline\hline
    \end{tabular}
  \end{center}
  \vspace{-7mm}
\end{table*}

The last two channels are intended to test the performance of the modes under mobility conditions, as this is a key aspect in wireless communications. Basically, mobility means rises and drops in the signal power received by the smartphone. Those variations translate into channel capacity fluctuations over time, a behavior that is fully captured by the two scenarios represented by the above-mentioned channels.

We have generated more representations than in previous experiments to increase the granularity and therefore enable a finer adaptation to the changing capacity of the channel to provide more accurate results and a clearer analysis. Thus, the original content was HEVC-encoded to produce 10 representations with different associated resolution and bit rate. The resulting values are presented in Table~\ref{tab:quality_ladder_experiments}. The tests simulate an ABR session of 3 hours. So, the content is played 15 times and there are 360 segments.

To illustrate the control policy of the client depending on the energy-saving strategies, Figure~\ref{fig:rep_energy_strategy} depicts one instance of a channel with a random behavior. On the left, it shows the available bandwidth throughout the session and the representation that is requested and downloaded, and later on decoded and presented to the user for the different energy-saving modes. On the right, it presents the energy consumed for the different strategies as a percentage of the one consumed if the energy-saving mode is off.

Table~\ref{tab:impact_strategies_on_quality_energy} reflects the impact of the proposed modes on the average image quality presented to the user and the average energy consumed as a percentage of the one consumed if the best representation possible is always selected. It includes both the absolute and differential values to facilitate the inspection. As expected, the energy saved if the light strategy is implemented is around 10\%, the savings obtained for the medium mode are about 20\% and the ones in the case of the strict strategy are close to 30\%. The reason why they do not reach the target percentages on average is twofold. The first one has to do with the limited number of qualities in the quality ladder. This leads to situations where, when the available bandwidth varies, the representation that is requested might be not that close to the targeted bit rate for the strategy in operation, with the corresponding impact on energy and quality. As the granularity increases, this effect is reduced. On the other hand, as the bandwidth goes down, the number of representation that can be selected is reduced, and thus all the strategies tend to request more similar bit rates, regardless of their original targets. This effect exacerbates when the bandwidth is particularly low and only the lowest representation is eligible. In this case, all the strategies cannot be implemented to any extent and the resulting energy consumption and video quality equal those of the off-mode. This effect can be seen several times along the session shown in Figure~\ref{fig:rep_energy_strategy}.

Regarding the average quality of the video provided to the user, as expected, we can see that it decreases as the energy-saving mode becomes more severe. However, it does not drop in the same degree for the different channel behaviors. Differences are more pronounced as the capacity of the channel decrease. In this sense, as a reference, it is worth noticing that VMAF differences ($\Delta$VMAF) below 6 points are typically not perceived by most viewers~\cite{vmafblog}. These results agree with those of previous studies~\cite{camara2019perceptually} and indicate that it is more advantageous to apply more severe energy-saving modes when the capacity of the channel is high, as it will impact only slightly the quality of the video presented to the users. As the available bandwidth decreases, switching to a worse quality has a greater impact on the video quality and so it is advisable to apply lighter modes.

\section{Conclusion}
\label{sec:conclusion}
This work presents a lightweight, non-intrusive segment quality selection policy aiming at extending battery lifetime of smartphone during ABR streaming sessions. The proposed technology, which is device- and connection-independent, can potentially benefit millions of consumers of video streaming applications in mobility scenarios, who could see the consumption of energy of their consumer device significantly reduced.

In particular, we propose several modes resulting from various trade-offs between energy consumption saving and video quality reduction. These modes derive from an exhaustive study analyzing and weighting the impact of several features (device, network connection, bit rate, video resolution...) on the energy spent by the device and the subsequent modeling of the energy consumption as a function of the relationship between the varying channel capacity and the bit rate of the representations in the quality ladder, as such relationship was found to impact highly on the amount of energy spent by the device along the session. The values that differed the most from said generalized model were those related to the 5G connection, a situation caused by the lack of DRX in the receiver for this type of wireless connection and that is expected to change when this mode is finally implemented in all commercial handheld devices.

Experiment results show the benefits of implementing such the proposed policy, whose exact outcome figures rely on the targeted energy saving and the QoE drop that can be tolerated.

\ifCLASSOPTIONcaptionsoff
  \newpage
\fi

\bibliographystyle{IEEEtran}
\bibliography{abr_energy_consumption_ieee_tce_2020_abbr}

\begin{IEEEbiography}[{\includegraphics[width=1in,height=1.25in,clip,keepaspectratio]{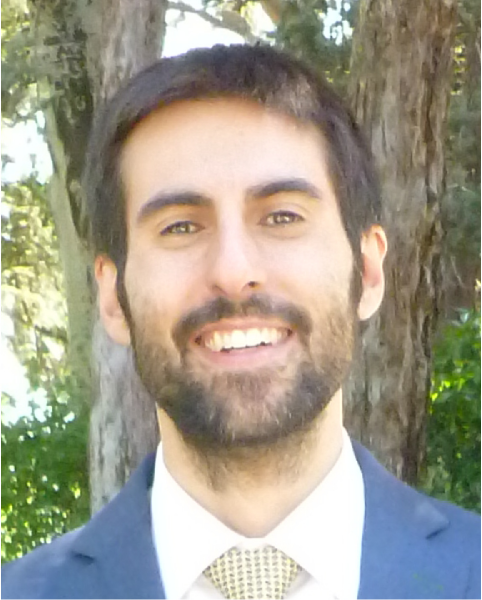}}]{C\'esar D\'iaz} received the Telecommunication Engineering degree (integrated BSc-MS) in 2007 and the Ph.D. degree in Telecommunication Engineering in 2017, both from the Universidad Polit\'ecnica de Madrid (UPM), Madrid, Spain. Since 2008 he has been a member of the Grupo de Tratamiento de Im\'agenes (Image Processing Group) of the UPM, where he has been actively involved in Spanish and European projects. His research interests are in the area of multimedia delivery and immersive communications.
\end{IEEEbiography}

\begin{IEEEbiography}[{\includegraphics[width=1in,height=1.25in,clip,keepaspectratio]{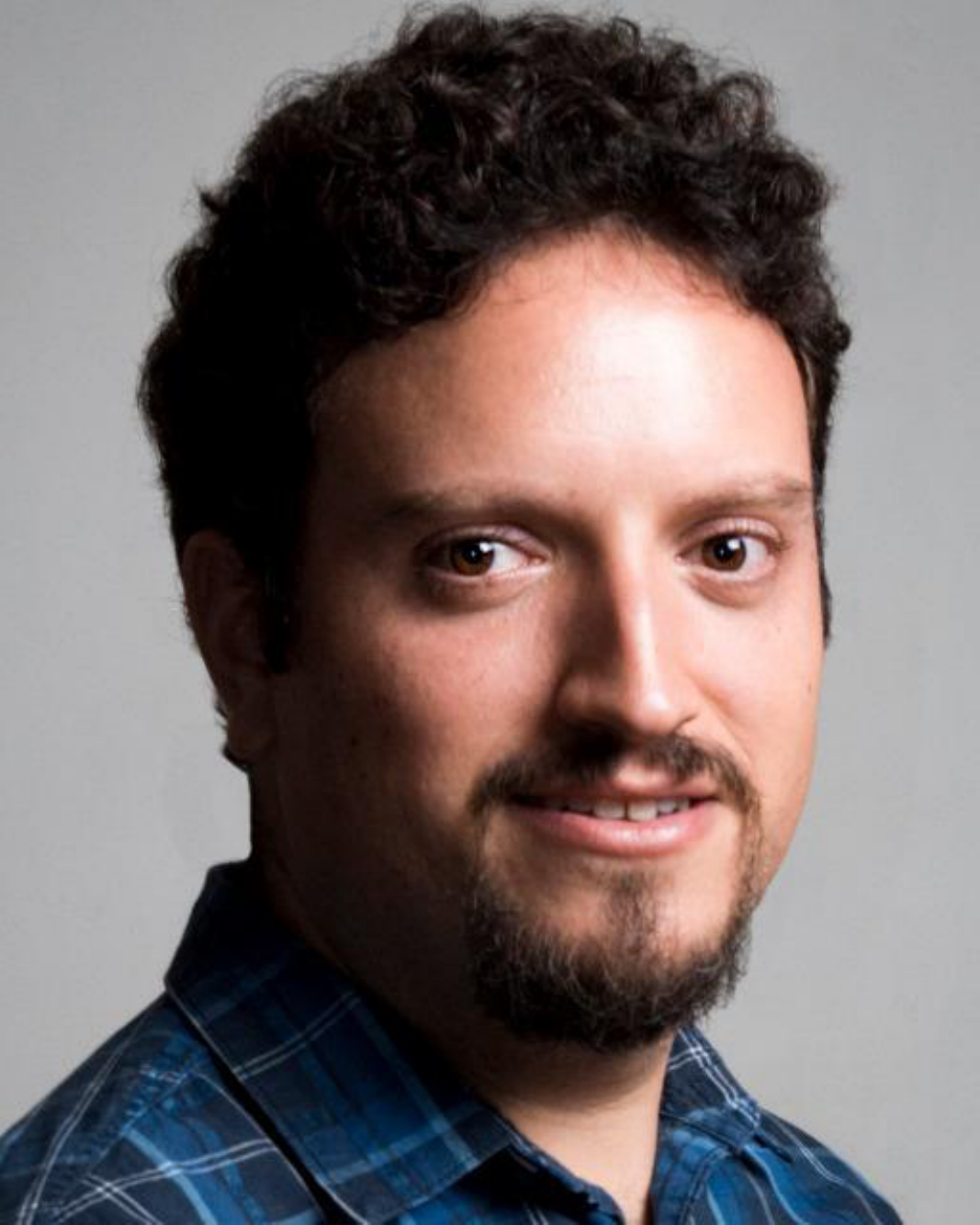}}]{Antonio Fern\'andez} received the BSc in Engineering in Telecommunication Technologies and Services in 2019 from the Universidad Polit\'ecnica de Madrid (UPM), Madrid, Spain. He has been a member of the Grupo de Tratamiento de Im\'agenes (Image Processing Group) of the UPM since 2017, where he has been actively involved in several research projects. His research interests are in the area of video capture, production and delivery.
\end{IEEEbiography}

\begin{IEEEbiography}[{\includegraphics[width=1in,height=1.25in,clip,keepaspectratio]{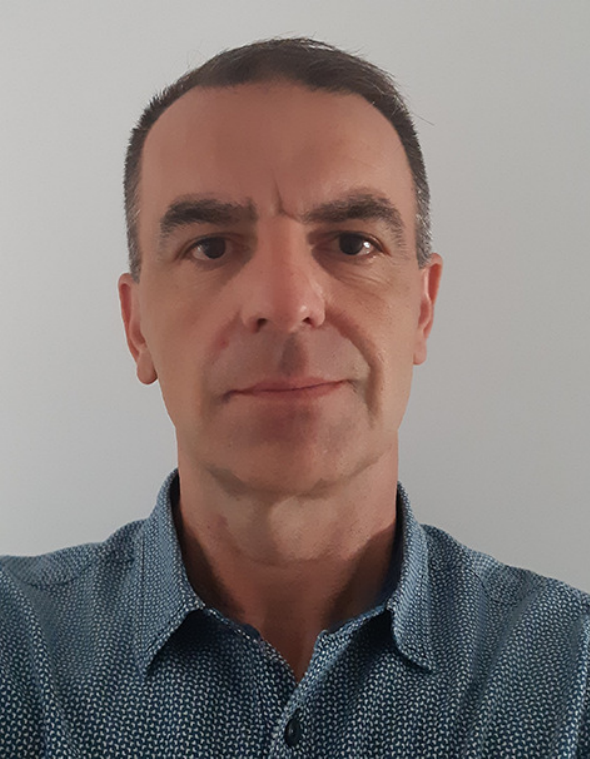}}]{Fernando Sacrist\'an} received the Computer Science degree from the Universidad Politécnica de Madrid (UPM) in 1998. He is a Video Applications Engineer at Nokia's Video Business Unit, where he is leading research activities focused on 5G and new video formats and services. He has also been an architect for the Intelligent Network Solutions and NextGen IN services at Nokia. He is an expert in video technology like encoding, format and transport, DRM, applications, as well as on the design, implementation and deployment of telecommunications services.
\end{IEEEbiography}

\begin{IEEEbiography}[{\includegraphics[width=1in,height=1.25in,clip,keepaspectratio]{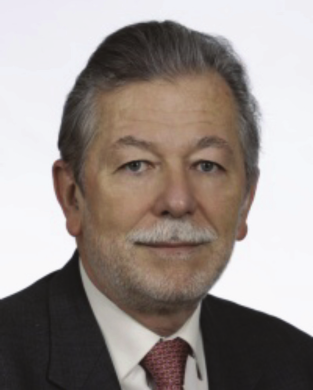}}]{Narciso Garc\'ia} received the Ingeniero de Telecomunicaci\'on degree (five years engineering program) in 1976 (Spanish National Graduation Award) and the Doctor Ingeniero de Telecomunicaci\'on degree (PhD in Communications) in 1983 (Doctoral Graduation Award), both from the Universidad Polit\'ecnica de Madrid (UPM), Madrid, Spain. Since 1977, he has been a member of the faculty of the UPM, where he is currently a Professor of Signal Theory and Communications. He leads the Grupo de Tratamiento de Im\'agenes (Image Processing Group), UPM. He has been actively involved in Spanish and European research projects, also serving as an evaluator, a reviewer, an auditor, and an observer of several research and development programs of the European Union. He was a Co-Writer of the EBU proposal, base of the ITU standard for digital transmission of TV at 34--45 Mb/s (ITU-T J.81). He was an Area Coordinator of the Spanish Evaluation Agency (ANEP) from 1990 to 1992 and he was the General Coordinator of the Spanish Commission for the Evaluation of the Research Activity (CNEAI) from 2011 to 2014. He has been the Vice-Rector for International Relations of the Universidad Polit\'ecnica de Madrid from 2014 to 2016. He was a recipient of the Junior and Senior Research Awards of the  Universidad Polit\'ecnica de Madrid in 1987 and 1994, respectively. His current research interests include digital video compression, computer vision, and quality of experience. 
\end{IEEEbiography}

\end{document}